\documentclass[12pt]{article}
\usepackage{amsmath,amssymb,amsfonts,amsthm}
\usepackage{graphicx}

\usepackage{epsfig,multicol}

\newcommand{\be}[0]{\begin{equation}}
\newcommand{\ee}[0]{\end{equation}}
\newcommand{\ba}[0]{\begin{eqnarray}}
\newcommand{\ea}[0]{\end{eqnarray}}

\title{Deep Inelastic Scattering off a Plasma with a Background Magnetic Field}

 \author{Majid Dehghani
\\
\footnotesize{ \textit{School of Physics,}}
\\
\footnotesize{ \textit{Institute for Research in Fundamental Sciences (IPM),}}
\\
\footnotesize{ \textit{P.O.Box 19395-5531, Tehran, Iran}}
\\ \footnotesize{e-mail: m.dehghani55@gmail.com } }
\begin{document}

 \date{\today}
\maketitle



 \newcommand{\norm}[1]{\left\Vert#1\right\Vert}
 \newcommand{\abs}[1]{\left\vert#1\right\vert}
 \newcommand{\set}[1]{\left\{#1\right\}}
 \newcommand{\R}{\mathbb R}
 \newcommand{\I}{\mathbb{I}}
 \newcommand{\C}{\mathbb C}
 \newcommand{\eps}{\varepsilon}
 \newcommand{\To}{\longrightarrow}
 \newcommand{\BX}{\mathbf{B}(X)}
 \newcommand{\HH}{\mathfrak{H}}
 \newcommand{\D}{\mathcal{D}}
 \newcommand{\N}{\mathcal{N}}
 \newcommand{\la}{\lambda}
 \newcommand{\af}{a^{ }_F}
 \newcommand{\afd}{a^\dag_F}
 \newcommand{\afy}{a^{ }_{F^{-1}}}
 \newcommand{\afdy}{a^\dag_{F^{-1}}}
 \newcommand{\fn}{\phi^{ }_n}
 \newcommand{\HD}{\hat{\mathcal{H}}}
 \newcommand{\HDD}{\mathcal{H}}

\abstract{Using holography, we analyse deep inelastic scattering of a flavor current from a strongly coupled quark-gluon plasma with a background magnetic field. The aim is to show how the magnetic field affects the partonic picture of the plasma. The flavored constituents of the plasma are described using D3-D7 brane model at finite temperature. We find that the presence of a background magnetic field makes it harder to detect the plasma constituents. Our calculations are in agreement with those calculated from other approaches. Besides the resulting changes for plasma structure functions a criteria will be obtained for the possibility of deep inelastic process in the presence of magnetic field.}


 \section{Introduction}\label{sec-intro}

 One of the main theoretical challenges in recent years is to describe the state of matter produced in a heavy ion collision, i.e. quark-gluon plasma (QGP). Experimental results indicate that the QGP may be in a strongly coupled regime.
 Many of the available formalisms in quantum field theory are perturbative, hence appropriate for a weakly coupled system.
  One needs non-perturbative techniques to handle strongly coupled systems and there are not too many tools for it. Lattice QCD has been the prime calculational tool for non-perturbative formulation of strong interacting systems. But it is of very limited use due to problems such as incorporation of chemical potential in the formalism. AdS/CFT correspondence (gauge/gravity duality), coming from string theory, is an interesting approach to describe strongly coupled systems \cite{maldacena1999large,Aharony:1999ti}. This conjecture relates a large-N gauge theory with large 't Hooft coupling $\lambda= g^2N $ to a weakly coupled string theory. The original correspondence was between $N=4$ supersymmetric Yang-Mills (SYM) theory and string theory in $AdS_5\times S^5 $.
  In the finite temperature version of theory the  $N=4$ SYM theory at finite temperature is related to the string theory in $AdS_5\times S^5 $ with a black hole in the bulk \cite{Witten:1998qj}.

 Deep inelastic scattering (DIS) is an important and powerful tool to study parton structure of hadrons. In this process an energetic electron collides the hadron and the exchanged virtual photon probes the internal structure of the hadron. Here the absorbtion of virtual photon by plasma is an inelastic process and is used to reveal the plasma partonic structure. More specifically absorbtion of the virtual space-like current excites some degrees of freedom in the plasma and it is used to mimic a DIS process. In the finite temperature version of the gauge/gravity formalism i.e. $AdS$ with blackhole, the virtual photon couples to massless fields in the adjoint representation or partons associated with gluons.
  To incorporate quarks i.e. fields in the fundamental representation, one way is to insert $N_f$ D7-branes in the $AdS_5\times S^5 $ Schwarzschild background \cite{Karch:2002sh}. Then the dual theory in the probe limit ($N_f \ll N_c$) is the $N=2$ SYM theory with $N_f$ hypermultiplets of fundamental fields. In this D3/D7 brane model the virtual photon couples to the fields in the fundamental representation or partons associated with quarks. As in a common DIS experiment the space-like gauge field living in the world volume of D7-brane probes the plasma structure. This space-like current describes an inelastic scattering of a flavor current off a strongly coupled $N=2$ SYM plasma.
  There are instances where gravity side gives a more clear view of the process. For example while the ability of plasma to absorb all perturbations (at sufficiently high temperatures) is not obvious from gauge theory side, it is almost straightforward to be deduced from the gravitational point of view and the "blackness" of black hole.
 To describe different absorption mechanisms of a current into the plasma one needs to study different thermal phases of plasma on the gravity side. It is well known that three different phases exist in the D3-D7 brane model \cite{Mateos:2007vn} : the low temperature Minkowski, the high temperature black hole and the critical embedding. Only Minkowski embedding will be considered here. There is an induced horizon on the D7-brane in the black hole embedding as it reaches down the black hole horizon. It is not difficult to deduce from a gravitational point of view that the space-like flavor current living on D7-brane world volume will ultimately fall into the black hole. In the dual description this is seen as a current that fluctuates into partons and branches due to the interaction with the plasma, until it disappears completely in the plasma \cite{Iancu:2009py, Bayona:2009qe}.

 The D7-brane in the Minkowski embedding  stays outside the black hole and is separated from the horizon in the radial direction. Because the current on D7-brane can not fall into the black hole the space-like flavor current absorbtion is not obvious. The new mechanism introduced in \cite{Iancu:2009py}, for flavor current absorbtion by plasma in the Minkowski embedding is based on the existence of D7-brane world volume vector mesons. Then the flavor current will disappear into the plasma by resonant production of the space-like vector mesons.

 In this paper we will discuss the kinematical changes of space-like current in the presence of a magnetic field for low and high energy regimes of Minkowski embedding. Consistency of our results with other approaches will be stressed. The effect of a background magnetic field on D7-brane embedding and some of its implications was investigated in \cite{Erdmenger:2007bn, Filev:2007gb, Ali-Akbari:2013txa, Ali-Akbari:2015bha}. Also in \cite{Filev:2011mt, Ammon:2012qs} the supergravity background with a fully backreacted set of massless/massive flavor D7-branes, with and without temperature, coupled to a non-vanishing B field was constructed.

 An important issue that needs to be considered by introducing a magnetic field into the problem is the breaking of $SO(3)$ rotational symmetry. This is another situation where the gravity side helps gaining insight into the problem. We may expect an asymmetry in the results from field theory point of view. But in this case the sole effect of a background magnetic field is to change the D7-brane embedding which is a function of radial holographic coordinate only. So the dependence on magnetic background field direction does not enter the problem. For simplicity  we will
 consider the most symmetric scenario in which the gauge field propagates in the magnetic field direction with a transverse polarisation.

 First we obtain the dispersion relation for transverse vector mesons in the presence of a magnetic field by expanding the vector meson modes in terms of regular and normalizable modes \cite{Mateos:2007vn, CasalderreySolana:2010xh}. From the resulting dispersion curve, we will discuss the implications for the limiting velocity and rest mass of the vector meson in the plasma. We will also discuss how the obtained dispersion curve, affects the DIS process. Then following \cite{Iancu:2009py, Bayona:2009qe}, using potential analysis of the Schrodinger-like equation of vector mesons we will see that how the space-like current potential changes in the presence of magnetic field. Potential analysis for different regimes of inelastic scattering energy shows that  absorption of the flavor current by plasma becomes harder in the presence of a magnetic field. An analysis of the implications for the potential at low and high temperature phases will be performed and the agreement with other approaches is stressed. As the resulting equations in the presence of the background magnetic field can't be solved analytically, all of the equations will be solved numerically.
 The first step is to solve the D7-brane embedding numerically without any approximation, so subsequent calculations such as vector meson equations and potential analysis, are done in a general numerically solved approach.

 Now we repeat the arguments of reference \cite{Iancu:2009py} for the conditions on kinematics of space-like flavor current that needs to be meet in order to contribute the DIS process off the plasma. Then we will check the possibility to satisfy the conditions in the presence of a background magnetic field. As explained in \cite{Iancu:2009py}, the relevant kinematics is given by $\omega \gg Q \gg T$ with $Q^2 >0$ and moreover space-like currents can excite mesons when they have high enough energies and small virtualities such that the currents and mesons are in a nearly light-like regime.
 The smallness of virtuality is needed because there exists a potential barrier near Minkowski boundary for large space-like virtualities.
  The current can penetrate in the inner region of D7-brane and excite mesons, if $\omega \geq Q^3/T^2$. The dispersion curve of figure [\ref{Dispersion curve}] in the next section, and also potential analysis of section \ref{Schroding poten}, leads to the conclusion that it is harder for the current to meet the above mentioned conditions and contribute to the DIS process. Specifically for the above relation, dispersion curve of figure [\ref{Dispersion curve}] shows that higher energies are needed for the current to reach the light-like region which is of relevance for the DIS process discussed here.
 The harder DIS process in the presence of magnetic field is also consistent with the factorization picture  \cite{Hatta:2007he, Collins:1991ty, Catani:1990eg}. The factorization picture state that the photon fluctuation into dipole ($q\bar{q}$ pair) and the target hadron also is described by a collection of dipoles.
 Our results and also the previously obtained ones \cite{Erdmenger:2007bn, Filev:2007gb}, show that a magnetic field makes the $q\bar{q}$ pair more bounded (so smaller in size) and it can be infered from the formalism that smaller size dipoles result in smaller cross sections for the scattering events.
 Another condition for the possibility of DIS process is the existence of an infinite tower of equally spaced
 levels. It is necessary that the levels be finely spaced for fixed and large values of $k$. In fact as will be discussed in the next section,  this can imposes a bound on the strength of the magnetic field for which the DIS can occur in a strongly coupled plasma.

The organization of the paper is as follows: in section \textbf{\ref{D3D7 model}} the D3/D7 model with a background magnetic field on the D7-brane is introduced and the induced metric on the D7-brane will be obtained. The transverse vector meson equations of motion will be obtained as the second order fluctuations of the D7-brane. Using the equations of motion and expansion in normalizable  modes, the dispersion curve will be plotted, followed by discussions on the implications of magnetic field for these curves. Writing the equations for transverse vector mesons in a Schrodinger-like form, in section \textbf{\ref{Schroding poten}} an analysis for the potentials and resulting changes due to the presence of a background magnetic field will be performed.
The main result of this paper will be presented in section \textbf{\ref{struc func sec}} by discussing  the effect of magnetic field on structure functions of plasma and occurrence of a DIS process.


\section{ Mesons in D3/D7 model with a background magnetic field}\label{D3D7 model}


The gravitational dual of $N=4$ super Yang-Mills theory at finite temperature is the decoupling limit of $N_C$ black D3-branes, where the corresponding metric is of the form (see appendix (\ref{appen D7 embedd})) \cite{maldacena1999large, Aharony:1999ti, Witten:1998qj}:

\begin{equation}\label{10 dim rho}
  dS^2=\frac{1}{2} \left(\frac{u_0 \rho}{L} \right)^2 \left( -\frac{f^2}{\tilde{f}}dt^2 + \tilde{f} dx^2 \right)+ \frac{L^2}{\rho^2} \left(d\rho^2 +\rho^2 d\Omega_{5}^{2} \right)
\end{equation}
where:

\begin{equation}\label{thermal func}
  f(\rho)=1-\frac{1}{\rho^4},  \;\;\;\;\;\;\;   \tilde{f}(\rho)=1+\frac{1}{\rho^4}
\end{equation}
Then by introducing the following variables:

\begin{equation}\label{R variable}
  \rho^2= r^2+ R^2, \;\;\;\;\;   r=\rho cos\theta, \;\;\;\;\;  R=\rho sin\theta
\end{equation}
the induced metric on D7-brane becomes:

\begin{equation}\label{induced metric1}
  dS^2=\frac{1}{2} (\frac{u_0 \rho}{L})^2 \left(-\frac{f^2}{\tilde{f}}dt^2 + \tilde{f} dx^2 \right)+ \frac{L^2}{\rho^2} \left((1+\dot{R}^2)d\rho^2 +\rho^2 d\Omega_{3}^{2} \right)
\end{equation}
where $R(r)$ is the profile of D7 brane embedding. Now with a constant background magnetic field $(F_{ab}^{{\tiny BG}})$ the D7 brane action is of the form:

\begin{equation}\label{matrix M}
    M_{ab} \equiv P[g]_{ab}+ 2\pi \alpha' F_{ab}
\end{equation}
Using the induced metric obtained in appendix [\ref{appen induc metric}], the second order fluctuations of field strength is:

\begin{equation}\label{2D ord lagr}
 \sqrt{-det (M+2\pi \alpha' (\delta F))}= -\sqrt{-det(G)}Tr[(G^{-1}) \delta F (G^{-1}) \delta F ]
\end{equation}
 Now using the induced metric $G$ and the equations of motion for $R$ from the Lagrangian, one obtains the D7-brane embedding in the presence of the magnetic field. In Appendix \ref{appen D7 embedd} we have plotted the brane embedding with and without a magnetic field numerically. It can be seen that the magnetic filed causes the brane to be repelled away from the black hole horizon. We will see that the results obtained from other approaches, are consistent with the implications of repelled D7-brane embedding.
 The DBI action, quadratic in the gauge field is of the form:

\begin{equation}\label{DBI action}
  S_{D7}=-\frac{(2\pi \textit{l}_s^2)^2}{4} T_{D7}N_f \int d^8 \sigma \sqrt{-G} G^{mp}G^{nq} F_{mn}F_{pq}
\end{equation}
with metric $G$, the induced metric on D7-brane with a background magnetic filed, defined in Eq.(\ref{G metric}).
The equations of motion resulting from Lagrangian (\ref{DBI action}) is:

\begin{equation}\label{EOM field}
  \partial_m ( \sqrt{-G} G^{mp}G^{nq} F_{pq}) =0
\end{equation}
Here  plane wave solutions  will be considered:

\begin{equation}\label{plane wave}
  A_{\mu}(x,r)= A_{\mu}(r) e^{-i\omega t+ ikz}
\end{equation}
($\mu=t, x, y, z$),and the boundary condition for gauge field is of the form: $A_{\mu} (r \rightarrow \infty)=A_{\mu}^{(0)} $.
The profile of D7-brane embedding is obtained numerically from the equation of motion (\ref{R embedd eq}) and then substituting in the gauge field equations obtained from DBI action (\ref{EOM field}). Here only the transverse fields will be consider $i=x,y$, so the relevant field strength components are: $F_{ri}=\partial_r A_i$, $F_{ti}=\partial_t A_i \rightarrow -i\omega A_i  $, and $F_{zi}=\partial_z A_i \rightarrow ik A_i  $. Then the resulting equations of motion for transverse gauge fields are:

 \begin{equation}\label{Ai eqs}
  \ddot{A}_{i}+ \left[\partial_r ln(\sqrt{-G} G^{rr}G^{ii} ) \right] A_{i} +  \frac{G^{zz}}{G^{rr}} \left( \frac{f^2}{\tilde{f}^2} \omega^2 - k^2 \right) A_{i} =0
\end{equation}
Note that the effect of background magnetic field is contained in the D7-brane embedding which enters the above equation through the term $\sqrt{-G}$. The non-normalizable modes of Eq.(\ref{Ai eqs}) are used to calculate the structure functions of plasma in the next section. The normalizable solutions of Eq.(\ref{Ai eqs})  describe the vector mesons. Here we consider the normalizable modes to see the effect of the magnetic field on meson spectrum.

\begin{figure}
\centering
  \includegraphics[width=8cm]{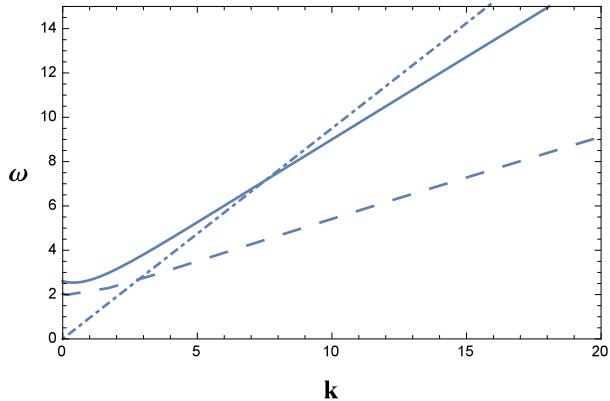}\\
  \caption{\footnotesize  Dispersion curve with (solid) and without (dashed) background magnetic field. The Dot-Dashed straight line correspond to $\omega=\upsilon k $ with $\upsilon < 1$. }\label{Dispersion curve}
\end{figure}

With the numerically solved embedding equation (Appendix A) and Fourier transforming the boundary coordinates, we obtain an eigenvalue equation for normalizable modes of vector mesons.
Imposing conditions at origin and boundary, results in discrete set of energy eigenvalues. If we expand transverse modes in terms of regular and normalizable modes \cite{CasalderreySolana:2010xh, Mateos:2007vn}:

\begin{equation}\label{expansion gauge}
  A_i(\omega,k,r)=\sum_n A_n(\omega,k)\xi_n(k,r)
\end{equation}
These normalizable modes with eigenvalues $\omega=\omega_n(k)$, obey the following equation:

\begin{eqnarray}\label{transv EOM2}
   \partial_r^2 \xi_n(k,r) + \left[\partial_r ln(\sqrt{-G} G^{rr}G^{ii} ) \right] \partial_r \xi_n(k,r)
    +  \left(\frac{G^{zz}}{G^{rr}} \right) k^2 \xi_n(k,r)                                                   \nonumber  \\
    =  \left( \frac{G^{zz}}{G^{rr}} \frac{\tilde{f}}{f} \right) \omega_n(k)^2 \xi_n(k,r)
\end{eqnarray}
Solving this eigenvalue equation for the cases with and without a background magnetic field, results in figure[\ref{Dispersion curve}]. Some important features of the transverse vector mesons in the presence of a magnetic field can be deduced from the figure, as explained below.
It can be seen from figure [\ref{Dispersion curve}] that increase in the slope results in longer path (higher energies) to reach light-like point of the dispersion curve, where it is  the region of concern here. Figure also shows that for asymptotically large $k$, the dispersion relation goes to a limiting velocity. Compared with the $B=0$ case it is evident that there is an increase in the limiting velocity due to increasing slope of the dispersion curve.
This result, again, is consistent with fact that the limiting velocity is the local velocity of light at the bottom of D7-brane and that the magnetic field repels the D7-brane toward the boundary, leads to a higher limiting velocity.

 Another effect related to the presence of a magnetic field can be obtained from expansion of dispersion relation in the following approximate form \cite{Iancu:2009py}:

\begin{equation}\label{dieper expan}
  \omega(k)\approx M_{rest} + \frac{k^2}{2M_{kin}},   \;\;\;\;\;\;\;\;\;\;  M_{rest}\approx M_{(0)}(1-2/R_0^4), \;\;\;   M_{kin}\approx \frac{M_{(0)}}{1-2/R_0^4}
\end{equation}
The dispersion curve intersection and slope of figure [\ref{Dispersion curve}] has been increased, so from the above formula we can conclude that the rest mass of the meson in the presence of a magnetic field has been increased. This is in agrement with the results of other works \cite{Erdmenger:2007bn, Filev:2007gb}. Their results show that mesons become more bounded in a background magnetic field.

Dispersion curve of figure \ref{Dispersion curve}, shows that increasing the momentum from zero drives the current to the light-like region of the meson which is the main region of concern in a DIS process. Comparing the curves with and without background magnetic fields shows that one needs more increase in momentum to reach the light-like region. So we see that more energy is needed to excite light-like mesons of D7-brane world volume. This result can be deduced following another line of reasoning \cite{Iancu:2009py}, where it imposes restrictions on the strongness of the magnetic field. The kinematics near light-like region and level spacing of meson spectrum is crucial for the DIS process to occur. The level spacing at fixed $k$ scales as $\delta\omega_n(k) \sim T(T/k)^{(1/3)}  $, and because the effect of background magnetic field is to increase the momentum needed for DIS process (to reach the light-like region), so increases the spacing. On the other hand the energy spacing $\Delta \omega \sim T(M_{gap}/T)^3 $ increases because of the increase in $M_{gap}$ . Then because the necessary condition for the DIS process to occur is: $\Delta \omega / \delta\omega_n \sim n^{1/3} (M_{gap}/T)^4 \gg 1$, so it can be concluded that the presence of magnetic field weakens this condition and a smaller possibility to detect partons of the plasma. Hence the magnetic field can be increased up to a level that the above relation holds. For higher values the conditions can't be satisfied and no partons can be detected.

\section {Schrodinger Potentials for Vector Mesons in a Magnetic field }\label{Schroding poten}

\begin{figure}
\begin{center}$
\begin{array}{cc}
\includegraphics[width=7cm]{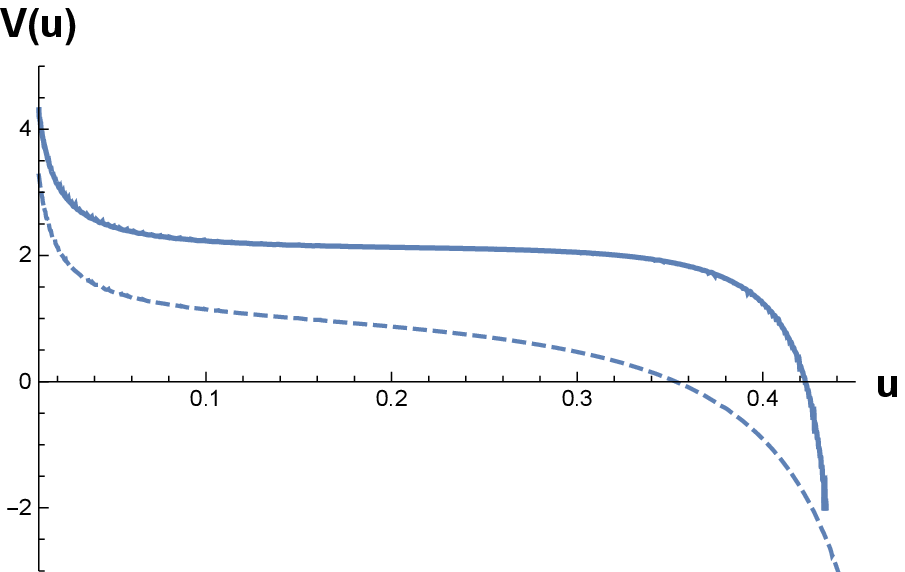} &
\includegraphics[width=7cm]{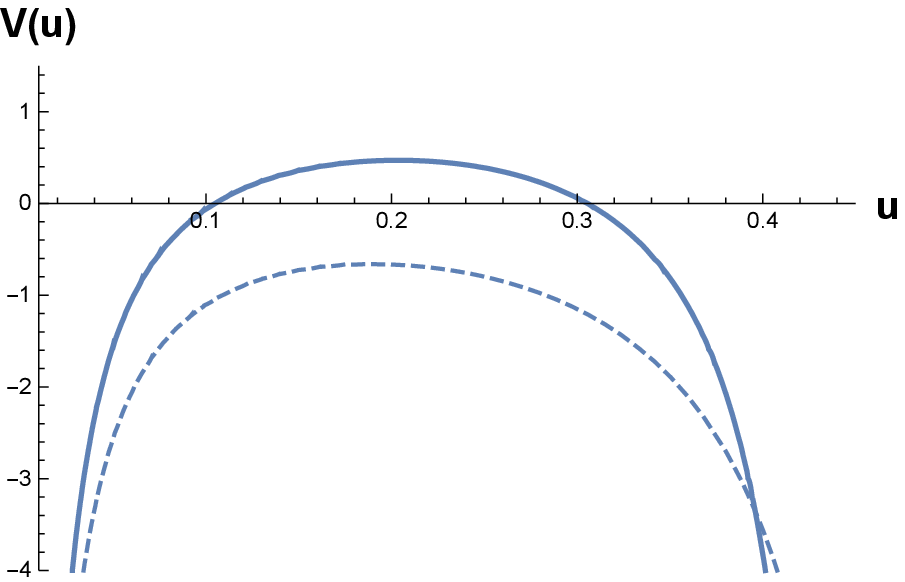}
\end{array}$
\end{center}
\caption{\footnotesize Low temperature potential energy for transverse flavor current, Left: for low temperature and $\omega^2 \ll Q^2$, Right: low energy and $Q^2 < \omega^2 $.   }\label{potential fig1}
\end{figure}

In this section we will discuss vector mesons of D7-brane in the Minkowski embedding i.e. low temperatures, with low and high energies. It will be explained how the magnetic field affects the gauge field potential. Then it will be possible to explain the effect of magnetic field on boundary perturbations and the way these perturbations will be absorbed by the
plasma. In order to write the equations of motion in a Schrodinger form and plotting the potential in a unit interval the following metric and D7-brane embedding will be used.
In the following metric the boundary is located at $u=0$ and black hole horizon at $u=1$  \cite{Bayona:2009qe}:

\begin{equation}\label{10 dim rho}
  dS^2= \frac{r_0^2 }{L^2 u} ( -(1-u^2)dt^2 +  d\vec{x}^2 )+ \frac{L^2}{4u^2(1-u^2)} du^2 +L^2 d\Omega_{5}^{2}
\end{equation}
In Appendix \ref{appen D7 embedd}, figure [\ref{D7 brane embdd}] right, we have plotted the effect of magnetic field on the D7-brane profile $\Theta(u)$ .  
For transverse gauge fields we use the gauge $A_u=0$, and the following plane wave ansatz:

 \begin{equation}\label{plane wave}
  A_{\mu}(x,u)=e^{-i\omega t + ikz}\bar{A}_{\mu}(u)
\end{equation}
If the brane profile $\Theta(u)$ is inserted into the quadratic DBI action equations of motion (\ref{Ai eqs}) and written in Schrodinger form:

\begin{equation}\label{Schrod form}
  \frac{d^2 A_i(u)}{du^2}- V(u)A_i(u)=0
\end{equation}
then one reads the potential for transverse gauge fields.
  From the dispersion curve of figure [\ref{Dispersion curve}], we see that the low temperature and sufficiently low momentum is in time-like region and increasing the energy drives the curve to light-like region. We have plotted the Schrodinger potential for low energies $\omega^2 \ll Q^2$ in figure [\ref{potential fig1}] left. As can be seen the magnetic field causes a thicker potential at the boundary, so harder for the flavor current to penetrate. Also it becomes an attractive potential at a larger distance from the boundary. These again, indicate the  conditions become more restricted to excite world-volume mesons.
 The plot in figure [\ref{potential fig1}] right, shows that at high energy and low virtuality the effect of magnetic field is to push the barrier towards the boundary and also squeezing the bottom of potential . Then this implies that smaller number of bound states can be formed, and in the light-like region of concern (where virtuality changes sign), smaller number of mesons will be available to absorb the flavor current energy.
 This conclusion can be deduced from the fact that the magnetic field repels the brane towards the boundary and the tip of brane gets closer to it.

\section {Plasma Structure functions in a Magnetic field }\label{struc func sec}

The system of $N_f$ D7 brane intersecting $N_c$ D3 black brane is an $N=2$ gauge theory with $N_f$ flavors
and global $U(N_f)\simeq SU(N_f)\times U(1)_q $ symmetry. The current $J_{q}^{\mu}$ associated with conservation of quark number corresponds to diagonal subgroup $U(1)_q $. To mimic the deep inelastic scattering process in this holographic set up, one introduces an abelian electromagnetic gauge field $A_{\mu}$ that minimally couples to $J_{q}^{\mu}$, i.e. $A_{\mu}$ acts as the source for current $J_{q}^{\mu}$. So the sketched process is the exchange of virtual photon ($A_{\mu}$) between the plasma and the propagating hard lepton.
The deep inelastic structure functions are obtained from the following current-current correlator:

\begin{equation}\label{curr curr}
  \Pi_{\mu\nu}(q)\equiv i \int d^4y e^{-iq.y} \theta(y_0)\langle[J_{q}^{\mu},J_{q}^{\nu}]\rangle_T
\end{equation}
where the $\langle...\rangle_T$ means thermal expectation value in the plasma. The structure functions are calculated from the imaginary part of real-time correlators as follows:

\begin{equation}\label{polar tensor}
  \Pi_1(x,Q^2)=-\frac{N_f N_c T^2}{8} \left[ r^3 \frac{\partial_r A_i(r,\omega,k)}{A_i(r,\omega,k)} \right]_{r\rightarrow \infty}
\end{equation}
and then DIS structure functions are obtained from:

\begin{equation}\label{struct func}
  F_1(x,Q^2)=\frac{1}{2\pi} Im \Pi_1,  \;\;\;\;\;\;\;\;\;    F_2(x,Q^2)\sim x F_1(x,Q^2)
\end{equation}

\begin{figure}
\centering
  \includegraphics[width=8cm]{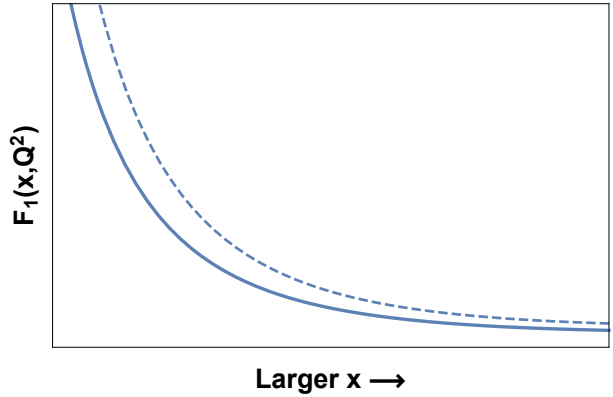}\\
  \caption{\footnotesize  Structure Functions for the cases with (Solid) and without (Dashed) background magnetic field $B$. }\label{strucureb fun}
\end{figure}

By inserting the non-normalizable modes of gauge fields in Eq.[\ref{polar tensor}] the structure function of plasma is calculated numerically for the cases with and without a background magnetic field. It can be seen from figure [\ref{strucureb fun}] that the magnetic field pushes the structure functions towards smaller values of Bjorken variable $x$. So the plasma starts to show partonic behavior in a DIS experiment, at smaller values of $x$. In order to interpret the resulting changes, we rewrite the functional form of the structure function obtained in \cite{Iancu:2009py, Bayona:2009qe}, as:

\begin{equation}\label{struc func result}
 F_1 (x,Q^2) \approx \frac{3}{4\Gamma^2(1/3)} N_f N_c T^2 \left( \frac{Q^2}{12\pi T^2x} \right)^{(2/3)}
\end{equation}
From the plots of figure [\ref{strucureb fun}], we can deduce that the magnetic field doesn't change the functional form of the structure function. If we relate the changes to the coefficient in the above equation, we can conclude that the background magnetic field effectively reduces the flavor degrees of freedom $N_f$ (the flavor current does't couple to gluon degrees of freedom, so $N_c$ unchanged).

\section {SUMMARY}\label{conclu}

In this work we studied the implications of a background magnetic field on a strongly coupled plasma. We showed that the presence of magnetic field causes the mesons in the plasma to become more bounded (heavier) and also the level spacing between them increased. These changes make the absorbtion of a flavored current harder and so less chance for DIS. The potential analysis of the Schrodinger-like equation and dispersion relations of mesons confirms the previous results. We showed that the structure function of the plasma will be pushed towards smaller $x$ values.
Also it was pointed out that the breaking of rotational symmetry by background magnetic field doesn't lead to anisotropic structure functions.
Possible next step is to study the resulting changes made on saturation line of the plasma phase by the background magnetic field. Also the case of black hole embedding where the magnetic field has a stabilising effect on it can be considered.

\section*{Acknowledgments}
I would like to thank Mohammad Ali-Akbari and Hajar Ebrahim for many useful discussions on the subject and numerics. The author thanks the hospitality of the School of Physics at Institute for Research in Fundamental Sciences (IPM).

\appendix
\numberwithin{equation}{section}
\addcontentsline{toc}{section}{Appendices}
\section*{Appendices}
\section { D7-brane embedding in deferent coordinates }\label{appen D7 embedd}

The dual of $N=4 SYM$ at finite temperature is the metric of the form:

\begin{equation}\label{10dim metric}
  dS^2= \frac{u^2}{L^2} \left(-f(u)dt^2 + dx^2 \right)+ \frac{L^2}{u^2} \left(\frac{du^2}{f(u)}+u^2 d\Omega_{5}^{2} \right)
\end{equation}
where $L=4\pi g_s N_c l_{s}^{4}$ is the curvature radius and $f(u)=1-u_{0}^{4}/u^4$, with $u_0=\pi L^2T$ the radial position of black hole horizon. Because we need to insert D7-branes in the above mentioned background, it is appropriate to perform the following change of radial coordinate:

\begin{equation}\label{radial coord}
  (u_0\rho)^2=u^2+\sqrt{u^4-u_0^4}
\end{equation}
With the above mentioned change of coordinate (\ref{10dim metric}) transforms to the metric of equation (\ref{10 dim rho}) in the text.
By the introduction of D7 brane in the above mentioned background we will have fields in the fundamental representation of $SU(N_c)$. The DBI action for a D7-brane with the given $AdS_5\times S^5$ metric (\ref{10 dim rho}), and a background gauge field is of the form:

\begin{equation}\label{induced plus magn}
  S_{D7}=\sqrt{det(P[g]+ 2\pi \alpha' F^{BG})}
\end{equation}
Then for a constant magnetic background, $F_{xy}^{BG}=B$ the action density of the D7 brane becoms:

\begin{equation}\label{D7 action}
  I_{D7}= \int d\rho r^3 f \sqrt{(\tilde{f}^2+B^2)(1+\dot{R}^2)}
\end{equation}
and the resulting equation of motion for $R(r)$:

\begin{equation}\label{R embedd eq}
  \partial_r \left[r^3 f \sqrt{\tilde{f}^2+B^2}\frac{\partial_rR}{\sqrt{1+(\partial_rR)^2}} \right]= r^3 \frac{\partial}{\partial R} \left(  f \sqrt{\tilde{f}^2+B^2}  \right) \sqrt{1+(\partial_rR)^2}
\end{equation}
This equation determines the shape of D7-brane embedding in a constant background magnetic field.
The boundaries of metric (\ref{10 dim rho}) corresponds to $\rho =1$ for black hole horizon and $\rho \rightarrow\infty$ for Minkowski boundary. In the potential analysis of the gauge fields on the D7-brane we need a coordinate system with boundaries in a finite interval. So the following metric is introduced \cite{Bayona:2009qe}:

\begin{equation}\label{compact 10D metric}
 ds^2= \frac{r_0^2}{L^2 u} [-(1-u^2)dt^2 + d\vec{x}^2 ] + \frac{L^2}{4u^2 (1-u^2)}du^2 + L^2 d\Omega_{5}^{2}
\end{equation}
where $r_0=\pi L^2 T$. The horizon is located at $u=1$ and the Minkowski boundary at $u=0$.
Here we also need to introduce D7-branes in the space of metric (\ref{compact 10D metric}). As in the previous case one needs to appropriately decompose the $S^5$ metric:

\begin{equation}\label{S5 decomp}
  d\Omega_{5}^{2}=d\Theta^2+ sin^2 \Theta d\Omega_{3}^{2} + cos^2 \Theta d\varphi^2
\end{equation}
Chooing $\varphi=0$ and $\Theta=\Theta(u)$ then the induced metric for D7-brane becomes:

\begin{equation}\label{induc D7 compa}
  dS^2= \frac{r_0^2 }{L^2 u} ( -(1-u^2)dt^2 +  d\vec{x}^2)+ L^2 \left( \frac{1}{4u^2(1-u^2)} + \Theta'^2 \right)du^2 +L^2 sin^2 \Theta d\Omega_{3}^{2}
\end{equation}

\begin{figure}
\begin{center}$
\begin{array}{cc}
\includegraphics[width=7cm]{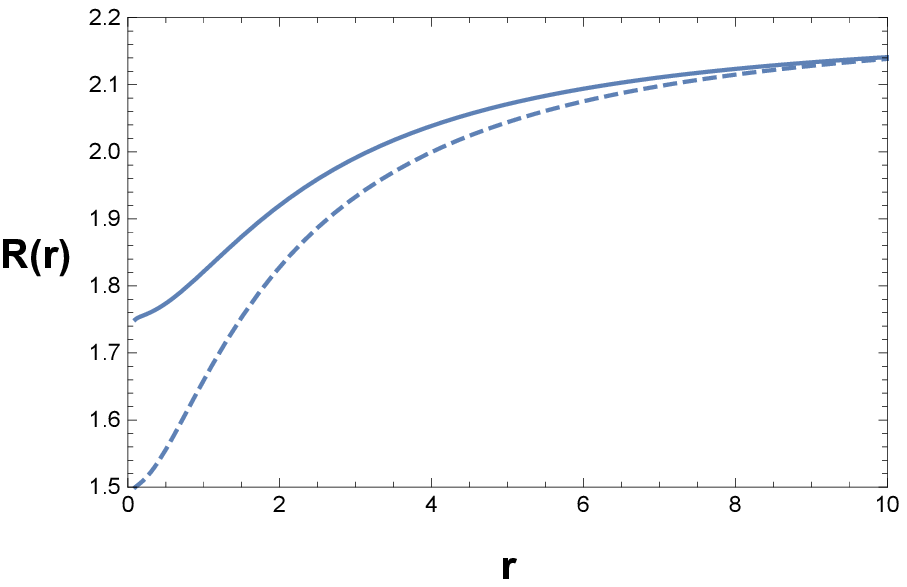} &
\includegraphics[width=7cm]{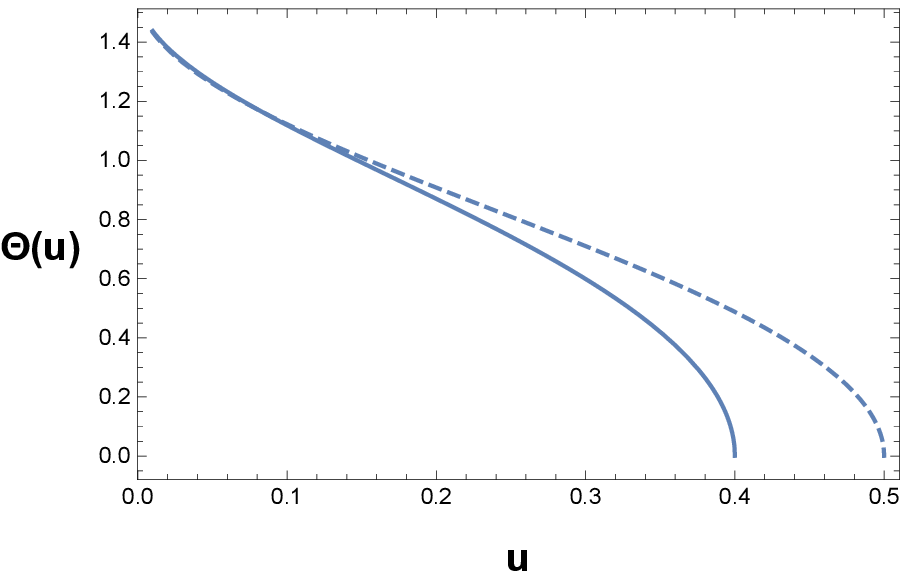}
\end{array}$
\end{center}
\caption{\footnotesize D7 brane embedding with (Solid) and without (Dashed) background magnetic field, Left: embedding for $R(r)$ obtained from Eq.(\ref{induced metric1}), Right: embedding for $\Theta(u)$ obtained from Eq.(\ref{induc D7 compa}) .   }\label{D7 brane embdd}
\end{figure}

Again by writing the DBI action in the form of Eq.(\ref{induced plus magn}) and the equations of motion for $\Theta (u)$, the D7-brane embedding in this metric in the presence of a background magnetic field will be obtained. Plots of the embedding with and without a magnetic field is shown in figure [\ref{D7 brane embdd}].

\section {Induced metric of D7-brane}\label{appen induc metric}

In this appendix, following \cite{Hashimoto:2013mua}, the induced metric on the D7-brane in the presence of a background magnetic field is investigated. The induced metric on D7-brane includes original induced metric $P[g]_{ab}$ and the background $F_{ab}$:

\begin{equation}\label{matrix M}
    M_{ab} \equiv P[g]_{ab}+ 2\pi \alpha' F_{ab}
\end{equation}
Denoting  the fluctuation field strength by $\delta F_{ab}$ and keeping $\delta F$ to second order, and decomposing the inverse metric into symmetric and anti-symmetric part, $M^{-1}\equiv (M^{-1})^s + (M^{-1})^a $, then we have:

\begin{equation}\label{determinant 1}
 \sqrt{-det (M+2\pi \alpha' (\delta F))}= -\sqrt{-det(M)}\times \frac{(2 \pi \alpha')^2 }{4} Tr[(M^{-1})^s \delta F (M^{-1})^s \delta F ]
\end{equation}
 the front factor decomposes to:

 \begin{equation}\label{front factor}
  \sqrt{-det(M)}= \sqrt{-det(((M^{-1})^s)^{-1})}\times \gamma , \;\;\;\;\;\;\; \gamma=det[1+((M^{-1})^s)^{-1}(M^{-1})^a]
\end{equation}
Now by introducing metric $G$:

\begin{equation}\label{G metric}
  G_{ab}= \mu_{7}^{1/2} \gamma^{1/2} (2\pi \alpha') \left[((M^{-1})^s)^{-1} \right]_{ab}
\end{equation}
The fluctuation Lagrangian takes the following form:

\begin{equation}\label{fluct lagra}
 \sqrt{-det (M+2\pi \alpha' (\delta F))}= -\sqrt{-det(G)}Tr[(G^{-1}) \delta F (G^{-1}) \delta F ]
\end{equation}


  \end{document}